\documentclass[a4paper,11pt]{article}
\usepackage{pos}
\usepackage{pgfplots}
\usepackage{relsize}
\usepackage{slashed}
\usepackage{wrapfig}
\usepackage{graphicx} 
\usepackage{subcaption}

\renewcommand{\logo}{\relax}

\title{Natural anomaly-mediation from the landscape with implications for LHC SUSY searches}
\author*[a]{Dibyashree Sengupta}
\affiliation[a]{INFN, Laboratori Nazionali di Frascati, Via E. Fermi 54, 00044 Frascati (RM), Italy}

\emailAdd{Dibyashree.Sengupta@lnf.infn.it}
\abstract{Supersymmetric models with the anomaly-mediated SUSY breaking (AMSB) have run into serious conflicts with 1. LHC \textit{sparticle} and Higgs mass constraints, 2. constraints from wino-like WIMP dark matter searches and 3. bounds from naturalness. These conflicts may be avoided by introducing changes to the underlying phenomenological models providing a setting for natural anomaly-mediation (nAMSB). We examine spectra of nAMSB arising from string landscape. Here, we investigated LHC constraints on nAMSB models that allow $m_{3/2}$ to lie within 90$-$200 TeV which may soon be discovered or falsified by a combination of 1. soft OS dilepton plus jet+ MET (OSDLJMET) searches which arise from higgsino pair production, 2. non-boosted hadronically decaying wino pair production searches and 3. same-sign diboson + MET searches arising from wino pair production followed by wino decay to W +higgsino. Some excess above SM background in the OSDLJMET channel already seems to be present in both ATLAS and CMS data.}

\FullConference{42nd International Conference on High Energy Physics (ICHEP2024)\\
18-24 July 2024\\
Prague, Czech Republic\\}

\begin{document}
\maketitle

\setlength{\abovedisplayskip}{3pt}
\setlength{\belowdisplayskip}{3pt}

\section{Introduction}

Anomaly-mediated SUSY breaking (AMSB) models originate from two different underlying set-ups. One of these, labeled as AMSB0, was given by Giudice et al.~\cite{Giudice:1998xp} and the other being the Randall-Sundrum ~\cite{Randall:1998uk} set-up, labeled as AMSB(RS).

\subsection{Minimal AMSB}

Ref.~\cite{Gherghetta:1999sw} and ~\cite{Feng:1999hg} postulated a minimal phenomenological AMSB (mAMSB) model with parameter space:
\begin{equation*}
    m_0,\ m_{3/2},\ tan\beta,\ sign(\mu)\qquad (mAMSB)
\end{equation*}
where $m_0$ is the additional common bulk term in the scalar masses. This phenomenological model appeals to both AMSB0 and AMSB(RS) model. However, current experimental limits on masses of \textit{sparticles} have rendered this model unnatural even in the context of the most conservative measure of naturalness $\Delta_{EW}$~\cite{Baer:2013gva}. Moreover, due to small values of the trilinear (A) terms, arising only from loop contribution, the mass of higgs obtained is much lower than its experimentally-obtained value $\sim 125$ GeV unless the top squarks are \textit{unnaturally} massive. Furthermore, this model yields wino-like dark matter~\cite{Moroi:1999zb} as \textit{wino} is the LSP. But pure wino-like dark matter has been excluded from WIMP search experiments~\cite{Cohen:2013ama}. 

\subsection{Natural AMSB}

These drawbacks of the mAMSB model has been addresed in Ref.~\cite{Baer:2018hwa} by adding a bulk term to the trilinear (A)-terms and adding a different bulk term to the masses of up and down type Higgs instead of the universal bulk scalar term. These changes in mAMSB formed a Natural AMSB (nAMSB) model with parameter space:
\begin{equation*}
    m_0(i),\ m_{H_u},\ m_{H_d},\ m_{3/2},\ A_0,\ tan\beta \qquad (nAMSB')
\end{equation*}

While the bulk term in $A$, denoted by $A_0$ assisted in pulling $m_h$ high making it more and more close to 125 GeV, adding a bulk term to $m_{H_u}$ and $m_{H_d}$ different from the other scalars gave freedom to trade $m_{H_u}$ and $m_{H_d}$ for phenomenologically more important parameters: $\mu$ and $m_A$. Therefore the revised parameter space becomes: 
\begin{equation*}
    m_0(i),\ m_{3/2},\ A_0,\ tan\beta,\ \mu,\ m_A  \qquad (nAMSB)
\end{equation*}
With the freedom to choose $\mu$, one can easily obtain suitable parameter space points with $\Delta_{EW} < 30$. Therefore, nAMSB satisfies naturalness constraint as well as yield higgs of mass $\sim$ 125 GeV. Moreover, since suitable choice of $\mu$ to respect naturalness constraint require $\mu < 350$ GeV, the LSP in this model automatically becomes higgsino-like which are found to be under-abundant for such small values of $\mu$~\cite{Baer:2018rhs}. Such WIMPs along with axion forms the entire dark matter content in this model. 

%\begin{figure}
%  \begin{minipage}[b]{0.45\linewidth}
%    \centering
%    \includegraphics[width=.95\linewidth, height = 5cm]{mh_ew.png} 
%    \label{fig:mamsb}
%    \vspace{0.5ex}
%
%  \end{minipage}%%
%  \hfill
%  \begin{minipage}[b]{0.45\linewidth}
%    \centering
%    \includegraphics[width=.95\linewidth, height = 5cm]{mh_ew_namsb.png} 
%    \label{fig:namsb}
%    \vspace{0.5ex}
      
%  \end{minipage} 
%  \caption{$\Delta_{EW}$ vs $m_h$ in (a) Minimal and (b) Natural AMSB Model}
%\end{figure}

%A typical mass spectra of nAMSB model is shown in Fig. 2.

%\begin{figure}
%    \centering
%    \includegraphics[width=10cm]{bm.png}
%    \caption{A typical superparticle mass spectrum generated from natural generalized anomaly mediation (nAMSB)}
%    \label{fig:my_label}
%\end{figure}

\subsection{nAMSB from the Landscape}

The string theory landscape ~\cite{Bousso:2000xa, Susskind:2003kw} along with Weinberg's anthropic principle ~\cite{Weinberg:1987dv} is immensely successful in explaining the tiny value of the Cosmological Constant ($\Lambda_{CC}$) $\simeq 10^{-120} m_P^2$. Therefore, one might find it rational to use a similar reasoning to predict masses of \textit{sparticles}. A general consideration of string theory landscape imply a mild statistical draw towards large soft SUSY breaking terms which is expected to be limited by anthropic requirement for sustenance of life, as we see it. In Agrawal et al. ~\cite{Agrawal:1998xa} (ABDS), it was shown that if $m_{weak}$ varies within (0.5-5) $\times$ $m_{weak}^{OU}$ which is the value of weak scale in our universe (OU), then heavy nuclei can be formed and life, as in our universe, can be sustained. Considering $m_{weak} \approx 4 \times m_{weak}^{OU}$ yields the condition $\Delta_{EW} \sim 30$. Hence, in this article, a mild statistical draw on soft SUSY breaking (SSB) terms towards large values is applied with a constraint of $\Delta_{EW} \leq 30$ to find most probable masses of \textit{sparticles} favoured by Nature. 

With this framework, the rest of the article is arranged as follows. In Sec.\ref{sec:land}, we assume a simple power law drag on the SSB terms to predict masses of Higgs and other \textit{sparticles} in nAMSB model. In Sec.\ref{sec:bm}, we present nAMSB benchmark and models lines, and show the nAMSB parameter space allowed by experiments. In Sec.\ref{sec:exp}, we discuss the discovery prospects of nAMSB at the current and upcoming runs of the LHC. Finally in Sec.\ref{sec:con} we conclude. Further details on this work can be found in Ref.~\cite{Baer:2023ech}.
\section{Masses of \textit{sparticles} and Higgs in \textit{nAMSB} from the Landscape}
\label{sec:land}
\begin{figure}[ht] 
  \begin{minipage}[b]{0.5\linewidth}
    \centering
    \includegraphics[width=.7\linewidth]{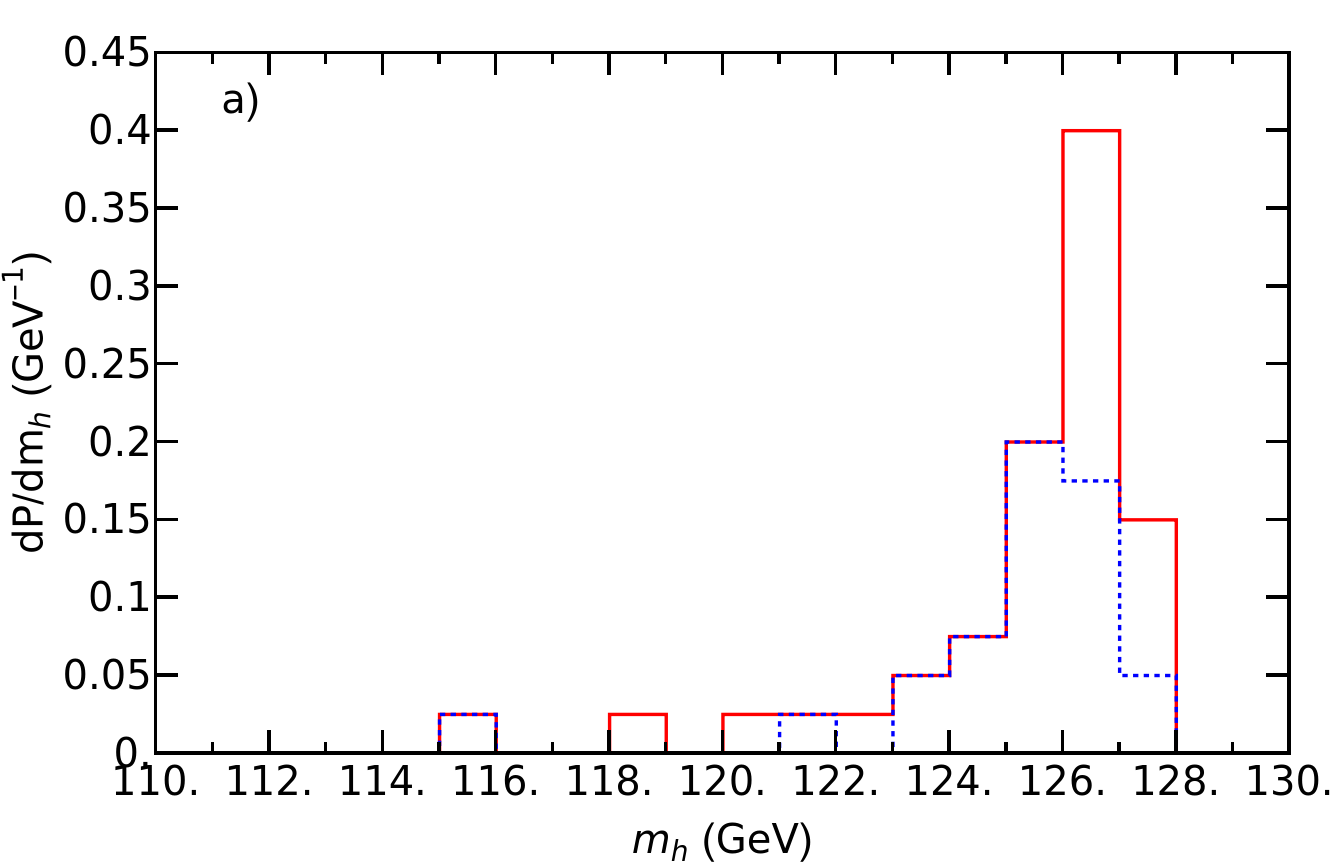} 
    %\vspace{0.5ex}
  \end{minipage}%%
  \begin{minipage}[b]{0.5\linewidth}
    \centering
    \includegraphics[width=.7\linewidth]{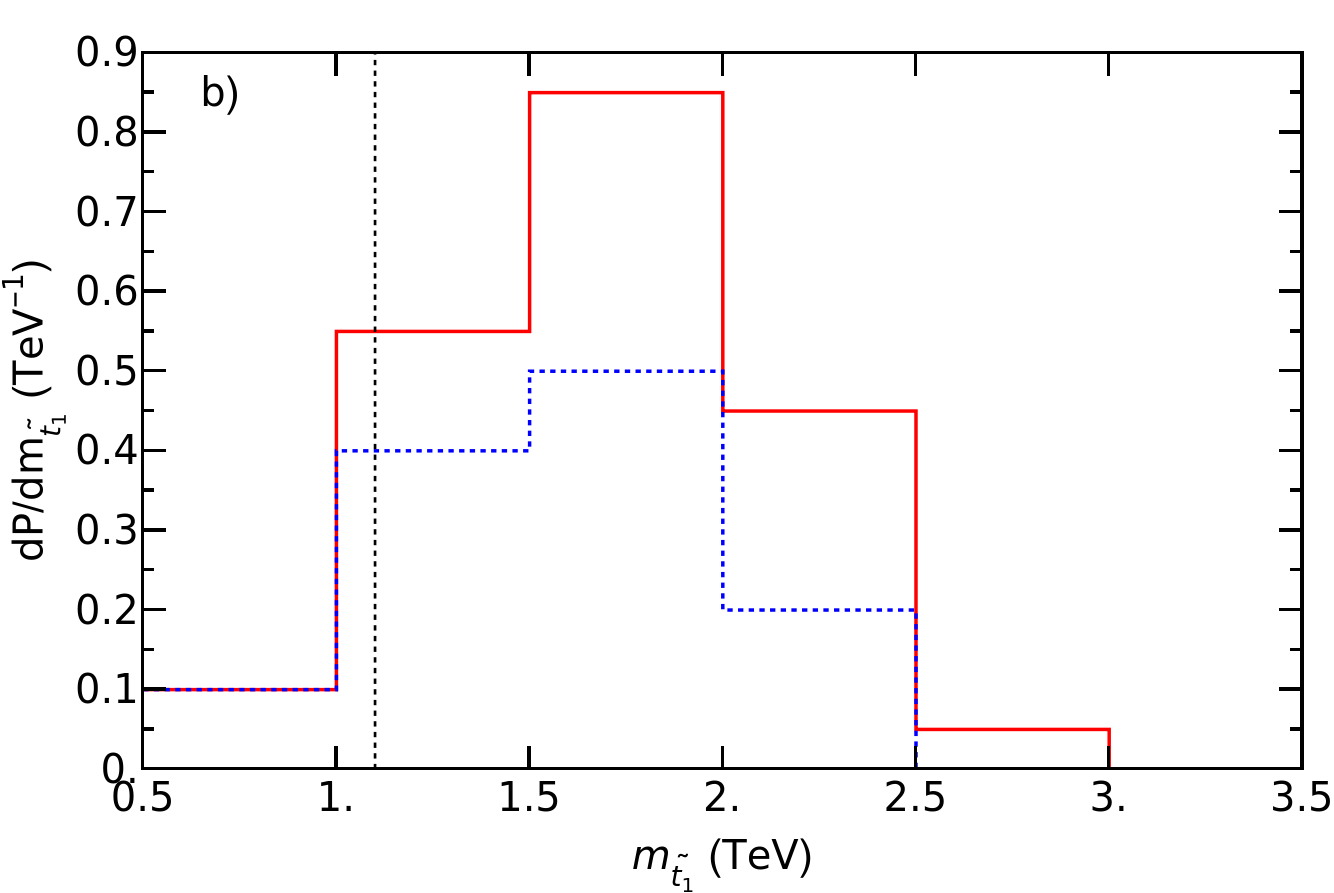}  
    %\vspace{0.5ex}
  \end{minipage} 
  \begin{minipage}[b]{0.5\linewidth}
    \centering
    \includegraphics[width=.7\linewidth]{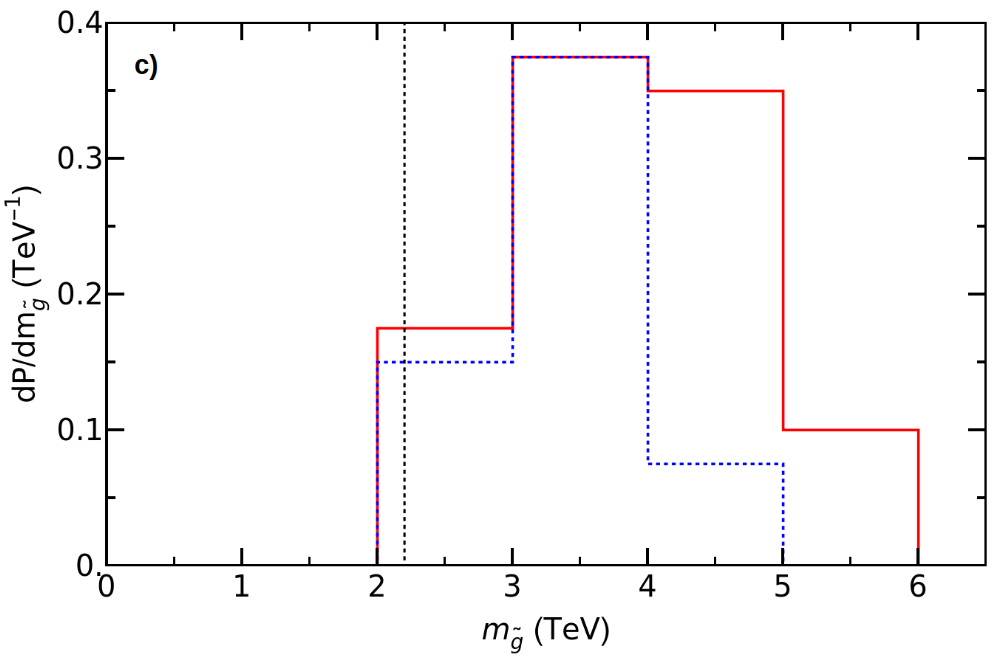} 
    %\vspace{0.5ex}
  \end{minipage}%% 
  \begin{minipage}[b]{0.5\linewidth}
    \centering
    \includegraphics[width=.7\linewidth]{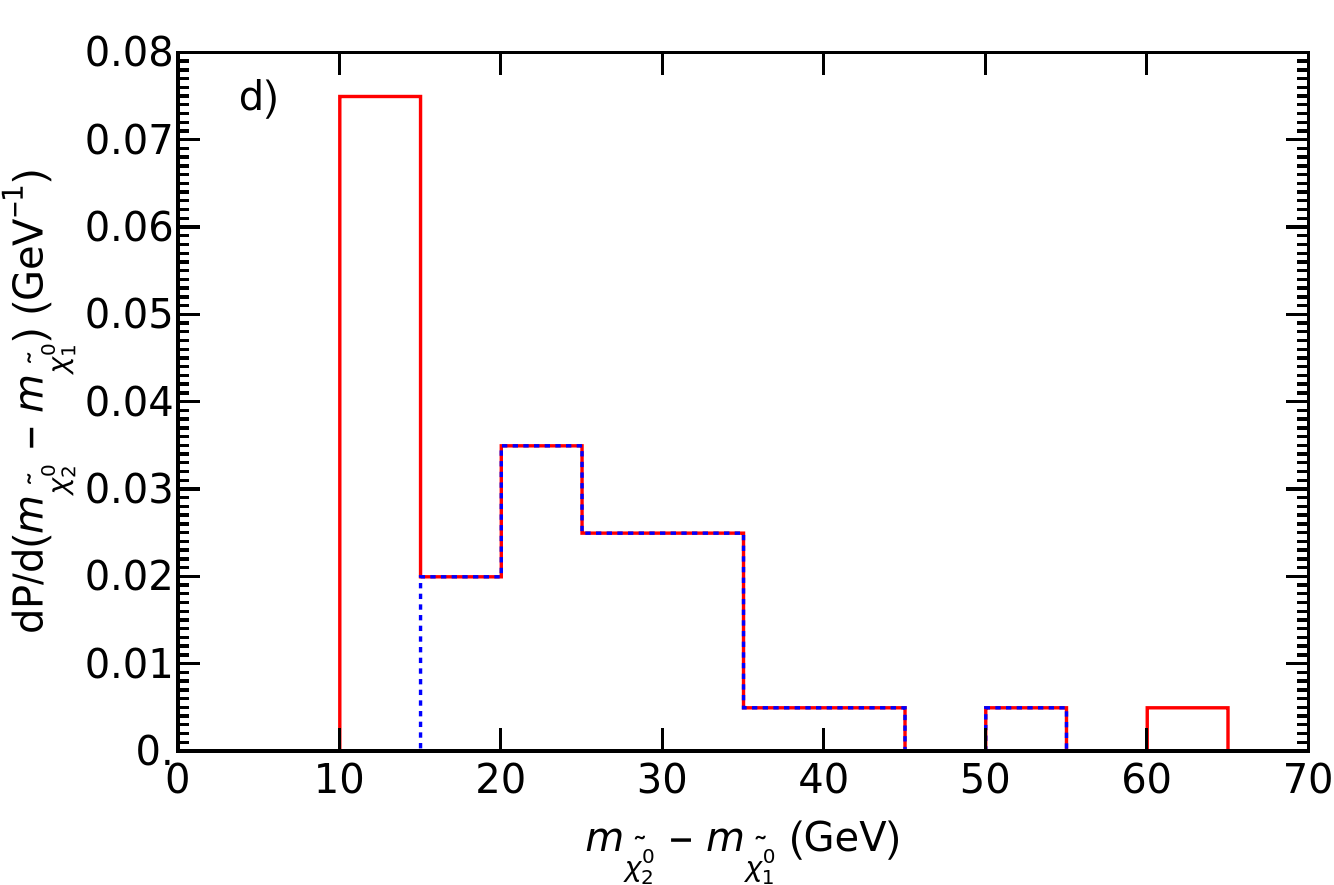} 
  \end{minipage} 
  \caption{Probability distribution of mass of (a) SM-Higgs (b) lightest stop (c) gluino (d) mass difference between two lightest neutralinos. }
  \label{fig:land}
\end{figure}

A power law drag of $n=1$ is applied to all the SSB terms in nAMSB model in the following range:
     $m_{3/2}$ : 80-400 TeV,
    $m_0(1,2)$ : 1-20 TeV,
     $m_0(3)$ : 1-10 TeV,
     $A_0$ : 0 - $\pm$20 TeV,
     $m_A$ : 0.25-10 TeV,
     tan $\beta$ : 3-60 (flat scan),
     $\mu$ = 250 GeV and the probability distribution of masses of Higgs and \textit{sparticles} are plotted in Fig~\ref{fig:land}. The red histogram shows the full probability distribution while the blue-dashed histogram shows the remaining distribution after LHC\textit{sparticle} mass limits are imposed. From Fig~\ref{fig:land}., one can infer that the nature prefers the SM Higgs mass to be $\sim 125-127$ GeV, quite in agreement with experiments while the most probable masses of the \textit{sparticles} are well beyond the current reach of LHC, also in agreement with the non-observance of \textit{sparticles} at the LHC. 

%\section{nAMSB benchmark and model lines}

\section{LHC-allowed nAMSB parameter space}
\label{sec:bm}
\begin{figure} [h!]
\begin{subfigure}[h]{0.32\linewidth}
\includegraphics[width=\linewidth]{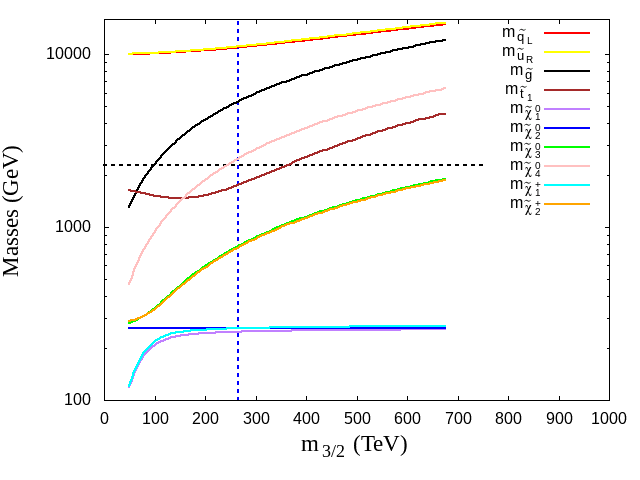}
\caption{\textcolor{red}{$m_{\tilde{g}}$ $\geq$ 2.3 TeV $\implies$ $m_{3/2}$ $\geq$ 90 TeV}; \textcolor{blue}{$\Delta_{EW}<30 \implies m_{3/2}$ $\leq$ 265 TeV}}
\label{bma}
\end{subfigure}
\hfill
\begin{subfigure}[h]{0.3\linewidth}
\includegraphics[width=\linewidth]{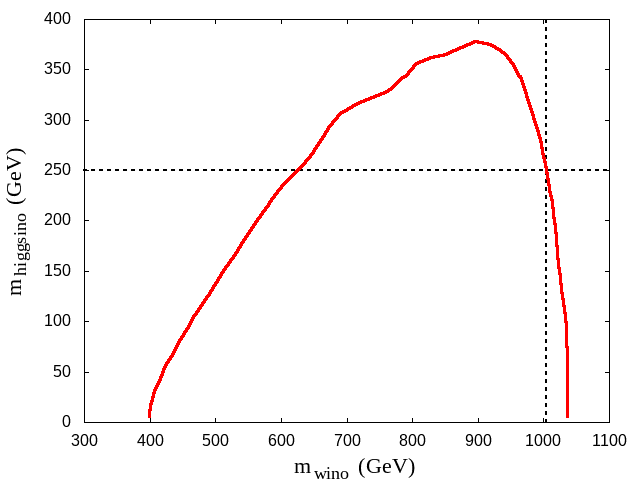}
\caption{$m_{wino}$ : 625 - 1000 GeV ($\sim$ \textcolor{red}{$m_{3/2}$ : 200 - 350 TeV}) ruled out}
\label{bmb}
\end{subfigure}
\hfill
\begin{subfigure}[h]{0.3\linewidth}
\includegraphics[width=\linewidth]{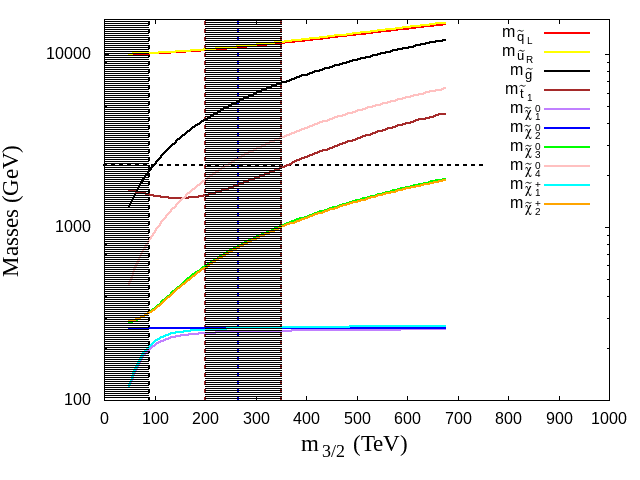}
\caption{nAMSB parameter space allowed/excluded by the LHC}
\label{bmc}
\end{subfigure}%
\caption{nAMSB model line: tan $\beta$ = 10, $m_0(1,2)$ = 10 TeV, $m_0(3)$ = 5 TeV, $A_0$ = 6 TeV}
\label{fig:bm}
\end{figure}

In nAMSB model, we fix the parameters to tan $\beta$ = 10, $m_0(1,2)$ = 10 TeV, $m_0(3)$ = 5 TeV, $A_0$ = 6 TeV and $m_{3/2}$ = 125 TeV and find a benchmark point which satisfy the naturalness constraint as well as all the experimental constraints. Next we elevate this benchmark point to a model line by keeping all the parameters fixed as above except for $m_{3/2}$ which is varied. Fig~\ref{bma}. shows the variation in\textit{sparticle} masses with $m_{3/2}$ for this model line. In Fig~\ref{bma}., the blue dashed line implies $m_{3/2} \leq 265$ TeV obtained from naturalness constraint i.e., requiring $\Delta_{EW} \leq 30$.  The black dashed line denotes the experimental limit on gluino searches $m_{\tilde{g}}$ $\geq$ 2.3 TeV which implies $m_{3/2} \geq 90$ TeV. 
\textbf{Constraints from $m_{wino}$:} Fig~\ref{bmb} shows the digitized ATLAS exclusion curve in the $m(wino) vs m(higgsino)$ plane obtained from recent ATLAS searches of EWino pair-production ~\cite{ATLAS:2021yqv}. A similar CMS search~\cite{CMS:2022sfi} has been performed in a smaller parameter space. The horizontal dashed line in Fig~\ref{bmb} shows our nAMSB model line. This plot implies that $m(wino): 625-1000$ GeV (which corresponds to  $m_{3/2}$ : 200 - 350 TeV) is ruled out. Therefore, experimental searches for winos limits $m_{3/2} \leq 200$ TeV. This limit will change with changing the higgsino mass parameter $\mu$ according to Fig~\ref{bmb}. Precisely for our nAMSB model the limits on $m_{3/2}$ are: \textbf{$m_{3/2}$: 90 - 200 TeV $\Longrightarrow$ Lower limit from $m_{\tilde{g}}$; Upper limit from $m_{wino}$ } as shown in Fig~\ref{bmc}. Therefore, an extremely small parameter space ($m_{3/2}$: 90 - 200 TeV) in nAMSB model, allowed by the LHC, is yet to be explored. Following this deduction, we discuss the implication of nAMSB model at the LHC in the next section.

%\begin{figure}
%    \centering
%    \includegraphics[width=9cm]{Mass.png}
%    \caption{Plot of\textit{sparticle} masses vs.$m_{3/2}$ along the \textcolor{purple}{nAMSB model line: tan $\beta$ = 10, $m_0(1,2)$ = 10 TeV, $m_0(3)$ = 5 TeV, $A_0$ = 6 TeV}. \textcolor{red}{Black dashed line: $m_{\tilde{g}}$ $\geq$ 2.3 TeV $\implies$ $m_{3/2}$ $\geq$ 90 TeV}; \textcolor{blue}{Blue dashed line: Upper limit on $m_{3/2}$ ($m_{3/2}$ $\leq$ 265 TeV) obtained from Naturalness($\Delta_{EW}<30$)}.}
%    \label{fig:my_label}
%\end{figure}    

%\begin{figure}
%    \centering
%    \includegraphics[width=9cm]{wino-higgsino.png}
%    \caption{Allowed/excluded regions of m(wino) vs. m(higgsino) plane from ATLAS analysis of EWino pair production followed by decay to W, Z, h with decay to boosted dijets. From this plot we would expect that the range $m_{wino}$ : 625 - 1000 GeV would be ruled out, corresponding to a range of \textcolor{red}{$m_{3/2}$ : 200 - 350 TeV}.}
%    \label{fig:my_label}
%\end{figure}

%\begin{figure}
%    \centering
%    \includegraphics[width=9cm]{Mass2.png}
%    \caption{Regions of the chosen nAMSB model-line along with various\textit{sparticle} masses allowed/excluded by the LHC.}
%    \label{fig:my_label}
%\end{figure}

\section{Prospects for nAMSB AT Run3 and HL-LHC searches}
\label{sec:exp}
\begin{figure} [h!]
\begin{subfigure}[h]{0.32\linewidth}
\includegraphics[width=\linewidth]{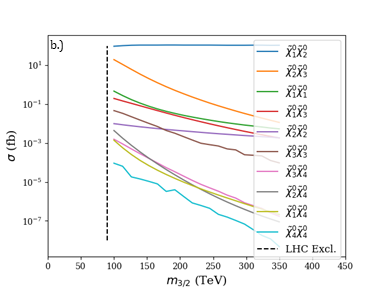}
\caption{neutralino pair production vs $m_{3/2}$}
\label{bra}
\end{subfigure}
\hfill
\begin{subfigure}[h]{0.3\linewidth}
\includegraphics[width=\linewidth]{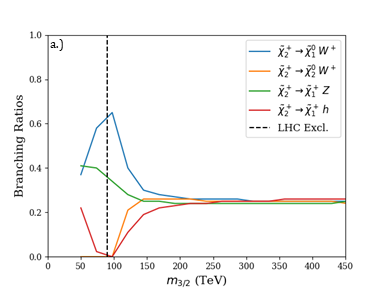}
\caption{BF($\tilde{\chi}_2^{+}$)}
\label{brb}
\end{subfigure}
\hfill
\begin{subfigure}[h]{0.3\linewidth}
\includegraphics[width=\linewidth]{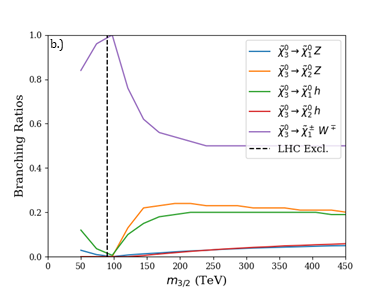}
\caption{BF($\tilde{\chi}_3^{0}$)}
\label{brc}
\end{subfigure}%
\caption{nAMSB model line: tan $\beta$ = 10, $m_0(1,2)$ = 10 TeV, $m_0(3)$ = 5 TeV, $A_0$ = 6 TeV}
\label{br}
\end{figure}

\textbf{OSDLJMET:} In nAMSB, just like any Natural SUSY model, the LSP ($\tilde{\chi}_1^0$) is higgsino-like and so is $\tilde{\chi}_2^0$ and $\tilde{\chi}_1^{\pm}$ with very small mass difference among them. Since $\tilde{\chi}_1^0$, $\tilde{\chi}_2^0$ and $\tilde{\chi}_1^{\pm}$ are the lightest \textit{sparticles}, pair production of higgsinos yields a huge cross-section as seen in Fig~\ref{bra}. However, owing to the small mass difference between the LSP($\tilde{\chi}_1^0$) and the heavier higgsinos($\tilde{\chi}_2^0$/$\tilde{\chi}_1^{\pm}$), the later decays to $\tilde{\chi}_1^0$ and very soft opposite-sign dilepton. To improve the visibility of such soft leptons, a high $p_T$ jet is used as a trigger. Therefore, \textbf{O}pposite-\textbf{S}ign \textbf{D}i\textbf{L}epton + high $p_T$ \textbf{J}et + $\slashed{E_T}$ (\textbf{MET}) from the LSP (\textbf{OSDLJMET}) proves to be a smoking gun signature at the LHC which will appear as excess in the plot of invariant mass of $\ell$ $\bar{\ell}$ at low values of $m_{\ell \bar{\ell}}$ equivalent to the small mass difference between the LSP and heavier higgsinos. 

\textbf{Wino pair-production:} In nAMSB model, winos are relatively lighter than in other Natural SUSY models. Hence signatures involving wino pair production can prove to be lucrative to look for nAMSB model. The branching fractions of the winos $\tilde{\chi_2}^{\pm}$ and $\tilde{\chi_3}^{0}$ are shown in Fig~\ref{brb} and \ref{brc} respectively which shows wino pair production followed by winos decaying to non-boosted jets + $\slashed{E_T}$ via $W^{\pm}$, $Z$, $H$ decaying hadronically in the allowed parameter space of nAMSB characterized by $m_{3/2}: 90-200$ TeV is a signature worth looking for. Fig~\ref{brb} and \ref{brc} also shows that the same-sign diboson (SSdB) + $\slashed{E_T}$ signature arising from wino pair production followed by winos decaying to $W^{\pm}$ and a higgsino also appears to be a smoking gun signature for wino searches at the LHC. 

%\begin{figure}
%\centering
%\begin{subfigure}{.55\textwidth}
%  \centering
%  \includegraphics[width=0.85\linewidth]{osdljmet.png}
%  \caption{Feynman diagram for opposite-sign dilepton+jets+MET signature from higgsino pair production at hadron colliders}
%  \label{fig:sub1}
%\end{subfigure}%
%\begin{subfigure}{.55\textwidth}
%  \centering
%  \includegraphics[width=0.85\linewidth]{zizj.png}
%  \caption{Plot for neutralino pair production vs $m_{3/2}$ along the nAMSB model line}
%  \label{fig:sub2}
%\end{subfigure}
%\end{figure}

%\begin{figure}
%\centering
%\begin{subfigure}{.55\textwidth}
%  \centering
%  \includegraphics[width=0.85\linewidth]{bfx2posm32.png}
%  \caption{}
%  \label{fig:sub1}
%\end{subfigure}%
%\begin{subfigure}{.55\textwidth}
%  \centering
%  \includegraphics[width=0.85\linewidth]{bfx30m32.png}
%  \caption{}
%  \label{fig:sub2}
%\end{subfigure}
%\caption{Plot of charged and neutral wino branching fractions a) BF($\tilde{\chi}_2^{+}$) and b) BF($\tilde{\chi}_3^{0}$)
%vs $m_{3/2}$ along the nAMSB model line}
%\label{fig:test}
%\end{figure}

\section{Conclusion}
\label{sec:con}
Although Minimal AMSB model has been rendered unnatural by the current experimental limits on the \textit{sparticles}, it can be modified to form a Natural AMSB model that satisfies naturalness constraint as well as all the limits from experiments. Given the success of string Landscape argument accompanied by anthropic argument in explaining the fine-tuning in $\Lambda_{CC}$, we are tempted to use the same to predict the masses of the SM Higgs and the \textit{sparticles} in the nAMSB framework and we found that the predictions satisfy the experimental constraints. Confronting the nAMSB model with naturalness constraint and all existing experimental limits provide us the small parameter space allowed:  $m_{3/2}: 90-200$ TeV. Lucrative channels to probe this allowed parameter space of nAMSB model at the current and upcoming runs of the LHC are: i) OSDLJMET from higgsino pair-production, ii) non-boosted jets + $\slashed{E_T}$ and SSdB + $\slashed{E_T}$ from wino pair-production. 

\acknowledgments{I thank all the organisers of the $42^{nd}$ International Conference on High Energy Physics (ICHEP) 2024 for their kind hospitality. I thank my collaborators Howard Baer, Vernon Barger, Jessica Bolich and Juhi Dutta. This research was supported by the U.S. Department of Energy, Office of Science, Office of High Energy Physics under Award Number DE-SC-000995 and William F. Vilas Estate.}

\bibliographystyle{JHEP}

\bibliography{reference}
\end{document}